\documentclass[final,5p,times,twocolumn]{elsarticle}
\usepackage{lineno,hyperref,amsmath,amsthm,amssymb,amsfonts,ragged2e,color,subfig}
\journal{``Plasma Physics Reports"}
\begin{document}
\begin{frontmatter}
\title{Dust-acoustic envelope solitons in an electron depleted plasma}
\author{J. Akter$^{*,1}$, N.A. Chowdhury$^{**,2}$, A. Mannan$^{\dag,1,3}$, and A.A. Mamun$^{\ddag,1,4}$}
\address{$^{1}$Department of Physics, Jahangirnagar University, Savar, Dhaka-1342, Bangladesh\\
$^{2}$Plasma Physics Division, Atomic Energy Centre, Dhaka-1000, Bangladesh\\
$^{3}$Institut f\"{u}r Mathematik, Martin Luther Universit\"{a}t Halle-Wittenberg, D-06099 Halle (Saale), Germany\\
$^{4}$Wazed Miah Science Research Centre, Jahangirnagar University, Savar, Dhaka-1342, Bangladesh\\
E-mail: $^{*}$akter277phy@gmail.com, $^{**}$nurealam1743phy@gmail.com\\
$^{\dag}$abdulmannan@juniv.edu,  $^{\ddag}$mamun\_phys@juniv.edu}
\begin{abstract}
A theoretical investigation of the modulational instability (MI) of dust-acoustic waves (DAWs)
by deriving a nonlinear Schr\"{o}dinger equation in an electron depleted
opposite polarity dusty plasma system containing non-extensive positive ions has been presented.
The conditions for MI of DAWs and formation of envelope solitons have been investigated.
The sub-extensivity and super-extensivity of positive ions are seen to change the
stable and unstable parametric regimes of DAWs. The addition of dust grains
causes to change the width of both bright and dark envelope solitons. The findings of this study may be helpful
to understand the nonlinear features of DAWs in Martian atmosphere, cometary tail, solar system,
and in laboratory experiments, etc.
\end{abstract}
\end{frontmatter}
\section{Introduction}
\label{1sec:Introduction}
The ubiquitous existence of the massive dust grains in space environments (viz., supernova explosion \cite{Sahu2012a},
Earth polar mesosphere \cite{Hossen2016a}, cometary tail \cite{Hossen2016b}, Martian atmosphere \cite{Hossen2016b},
interstellar clouds \cite{Hossen2017}, circumstellar clouds \cite{Hossen2017},
solar system \cite{Hossen2017}, and Earth's lower ionosphere \cite{Kopnin2007}, etc.) as well as  laser-matter
interaction \cite{Shahmansouri2013a} has been created a great interest among the plasma physicists \cite{Ferdousi2017,Ferdousi2015,Sahu2012b,Kopnin2009,Popel1994},
and is considered to develop new and modified low frequency electrostatic dust-acoustic (DA) waves (DAWs) \cite{Rao1990}
and dust-ion-acoustic waves (DIAWs) \cite{Shukla1992} and associated instabilities of these new eigen-modes.

The velocity distribution of energetic miniature particles (viz., ions and electrons) in space plasma is often
considered to follow the Maxwellian velocity distribution by assuming that these particles remain in
thermally equilibrium in the plasma system. But the concept of Tsallis statistics \cite{Tsallis1988}, first recognised by
Renyi \cite{Renyi1955}, has superseded the Maxwellian velocity distribution and rigorously applicable in explaining
the dynamics of the plasma medium which is not thermally equilibrium, and have also provided
the proof of existence of non-thermal particles which motion can not be described by Maxwellian
distribution function \cite{Saha2014,Zaghbeer2014a,Zaghbeer2014b,Roy2014,Bacha2012,Eastman1998}
in both space \cite{Pavlos2011} and laboratory experiments \cite{Liu1994}.
Tsallis \cite{Tsallis1988} introduced a parameter $q$, also known as non-extensive parameter, in his generalized
statistics in order to measure the degree of non-extensivity of the particular plasma system
containing particles moving very fast in comparison with their thermal velocities. Saha and
chatterjee \cite{Saha2014} considered a four component dusty plasma medium (DPM) comprising
opposite polarity dust grains (OPDGs) and non-extensive electrons and ions, and investigated
the nonlinear propagation of DA multi-soliton in the plasma medium, and found that the
values of $q$ control over the generation of compressive and rarefactive DA multi-soliton.
Zaghbeer \textit{et al.} \cite{Zaghbeer2014a} reported that the amplitude of the DA
shock waves increases with the increase of ion non-extensivity while decreases with the increase
of electron non-extensivity. Bacha and Tribeche \cite{Bacha2012} studied the
effects of electron non-extensivity on the potential structure of solitary waves in DPM,
and observed that the height of the pulse increases while the width of the pulse decreases
with increasing of $q$.

Modulational Instability (MI) is a well-known nonlinear mechanism in which the nonlinear self-interaction of the
carrier waves in the system causes to modulate the amplitude of the waves. Then, the evolution of
the system allows to localize the energy of the wave and is also governed by the nonlinear Schr\"{o}dinger equation (NLSE).
A number of authors \cite{Jahan2019,Saini2008,El-Taibany2006,Demiray2019,Moslem2011,Vladimirov1995,Gailitis1964,Gailitis1965,Vedenov1965}
have investigated the MI of the carrier waves in various plasma system with the help of NLSE in presence of massive dust grains (DGs).
Saini and Kourakis \cite{Saini2008} investigated the effects of dust concentration on the MI window of DAWs,
and found that an increment in the negative dust concentration leads to shrink the stable area of DAWs.
El-Taibany and Kourakis \cite{El-Taibany2006} reported that the addition of
non-thermal ions possibly stabilize the DA envelope soltion.
Demiray and Abdikian \cite{Demiray2019} performed a stability analysis in a warm DPM,
and observed that the MI occurs for small values of wave number and only bright envelope solitons
will propagate through the medium.

We assume here that electrons are significantly depleted during the charging of the negatively
charged dust grains. This makes $n_{i0}$, $Z_-n_{-0}$, $Z_+ n_{+0}>>n_{e0}$ valid. However, the minimum value of $n_{e0}/n_{i0}>m_e/m_i$
for $T_e\approx T_i$, where $m_e$ ($T_e$) is the electron mass (temperature).
The process of electron depletion has been identified by laboratory experiments and space observations,
and successively has been considered by many researchers \cite{Tagare1997,Ghai2018,Shahmansouri2012,Shahmansouri2013b,El-Tantawy2012}
to study the effects of the electron depletion in modifying the electrostatic potential structure in DPM.
Sahu and Tribeche \cite{Sahu2012a} examined the non-planar DA shock waves (DASHWs) in an
electron depleted DPM (EDDPM) in presence of non-extensive ions, and found that
the DA solitons may exhibit rarefaction and compression for $q>0$ and $q<0$, respectively.
Shahmansouri and Alinejad \cite{Shahmansouri2013a} analyzed DA solitary waves (DASWs) in a magnetized EDDPM.
Ghai \textit{et al.} \cite{Ghai2018} reported a theoretical analysis on the DASHW structure
regarding an electron depleted plasma system. Shahmansouri and Tribeche \cite{Shahmansouri2012}
studied the DA double-layers and DASWs in EDDPM having two temperature super-thermal ions.
Shahmansouri \cite{Shahmansouri2013b} theoretically analyzed the existence and dissipative nature
of DASWs in an unmagnetized EDDPM by considering dust-neutral collision.
Ferdousi \textit{et al.} \cite{Ferdousi2017} investigated DA shock excitations, and found
that the potential of the DA shock wave depends on the various physical parameters of the EDDPM.
In our article, we will observe the effects of electron depletion on the MI of DAWs and
associated DA envelope excitations in an unmagnetized EDDPM consisting of OPDGs and
$q$-distributed positive ions.

The article is arranged as follows: The governing equations describing our plasma
model are presented in section \ref{1sec:Governing Equations}. A standard NLSE has
been derived in section \ref{1sec:Derivation of the NLSE}. MI is given in
section \ref{1sec:Modulational instability}. The envelope solitons are presented in
section \ref{1sec:Envelope solitons}. A brief conclusion is finally provided in
section \ref{1sec:Conclusion}.
\section{Governing Equations}
\label{1sec:Governing Equations}
We consider a three-component EDDPM consisting of inertial negatively charged massive dust
grains (DGs), inertial positively charged DGs, and inertialess q-distributed positive
ions. At equilibrium, the charge neutrality condition for our considered plasma model
can be written as $Z_i n_{i0}+Z_+ n_{+0}\approx Z_- n_{-0}$, where $Z_i$, $Z_+$, and $Z_-$ are
the charge state of positive ions, positive and negative DGs, respectively, and
$n_{i0}$, $n_{+0}$, and $n_{-0}$ are the number densities of positive ions, positive
and negative DGs, respectively. Now, the the normalized form of the basic equations can be
written as
\begin{eqnarray}
&&\hspace*{-1.3cm}\frac{\partial n_{+}}{\partial t}+\frac{\partial}{\partial x}(n_{+} u_{+})=0,
\label{1eq:1}\\
&&\hspace*{-1.3cm}\frac{\partial u_{+}}{\partial t}+ u_{+}\frac{\partial u_{+}}{\partial x}=-\frac{\partial \phi}{\partial x},
\label{1eq:2}\\
&&\hspace*{-1.3cm}\frac{\partial n_{-}}{\partial t}+\frac{\partial}{\partial x}(n_{-} u_{-})=0,
\label{1eq:3}\\
&&\hspace*{-1.3cm}\frac{\partial u_{-}}{\partial t}+ u_{-}\frac{\partial u_{-}}{\partial x}=\nu_1\frac{\partial \phi}{\partial x},
\label{1eq:4}\\
&&\hspace*{-1.3cm}\frac{\partial^2\phi}{\partial x^2}=\nu_2n_--(\nu_2-1)n_i-n_+,
\label{1eq:5}\
\end{eqnarray}
where $n_i$, $n_{+}$, and $n_{-}$ are normalized by their equilibrium value $n_{i0}$,
$n_{+0}$, and $n_{-0}$, respectively; $u_+$ and $u_-$ are the positive and negative dust
fluid speed, respectively, normalized by positive DA speed $C_+=(Z_+k_BT_i/m_+)^{1/2}$,
(with $T_i$ being the positive ion temperature, $m_+$ being positive dust mass, and $k_B$
being the Boltzmann constant); $\phi$ represents the electrostatic wave potential normalized
by $k_BT_i/e$ (with $e$ being the magnitude of charge of an single electron); $t$ and $x$
are the time and space variables normalized by positive dust frequency $\omega_{P+}=(4\pi e^2Z_+^2n_{+0}/{m_+})^{1/2}$
and positive dust Debye length $\lambda_{D+} = (k_BT_i/4\pi e^2Z_+n_{+0})^{1/2}$, respectively;
Other parameters can be defined as $\nu_1=Z_-m_+/Z_+m_-$ and $\nu_2=Z_-n_{-0}/Z_+n_{+0}$.
It is important to note here that the mass of the negative DG is greater than
the mass of the positive DG (i.e., $m_->m_+$), the number density of the negative DG is
greater than the number density of the positive DG (i.e., $n_{-0}>n_{+0}$), and finally, the number of
charges residing on the negative dust particle is greater than the positive dust particle (i.e., $Z_->Z_+$).
Now, the expression for the number density of the non-extensive positive ions may be written as \cite{Saha2014}
\begin{eqnarray}
&&\hspace*{-1.3cm}n_i=[1-(q-1)\phi]^{\frac{q+1}{2(q-1)}}.
\label{1eq:6}\
\end{eqnarray}
The parameter $q$, given in the Eq. \eqref{1eq:6}, is a degree of measurement of non-extensivity
of plasma particles. When $q>1$, then the plasma system can be considered sub-extensive system while
the plasma system can be considered super-extensive system when $-1<q<1$, and  $q=1$ represents
a Maxwellian system. By substituting Eq. \eqref{1eq:6} into Eq. \eqref{1eq:5},
and expanding up to third order in $\phi$, we get
\begin{eqnarray}
&&\hspace*{-1.3cm}\frac{\partial^2\phi}{\partial x^2}+n_++\nu_2=1+\nu_2n_-+J_1\phi+J_2\phi^2+J_3\phi^3+\cdots,
\label{1eq:7}\
\end{eqnarray}
where
\begin{eqnarray}
&&\hspace*{-1.3cm}J_1=[(\nu_2-1)(q+1)]/2,
\nonumber\\
&&\hspace*{-1.3cm}J_2=[(1-\nu_2)(q+1)(3-q)]/8,
\nonumber\\
&&\hspace*{-1.3cm}J_3=[(1-\nu_2)(q+1)(3-q )(3q-5)]/48.
\nonumber\
\end{eqnarray}
We may note that the term on the right hand side is the contribution of $q$-distributed positive ions.
\begin{figure}
\centering
\includegraphics[width=80mm]{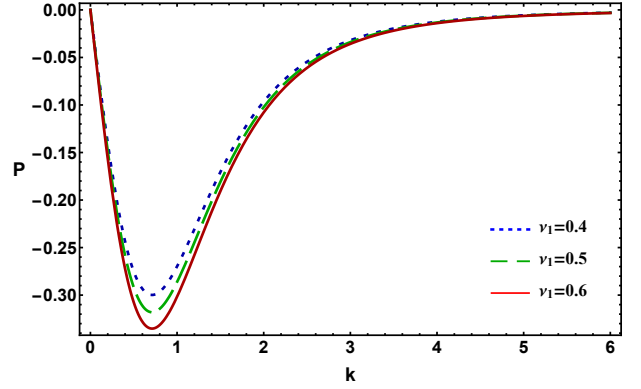}
\caption{Plot of $P$ vs $k$ for different values of $\nu_1$ when $\nu_2=2.5$ and $q=1.7$.}
\label{1Fig:F1}
\end{figure}
\section{Derivation of the NLSE}
\label{1sec:Derivation of the NLSE}
We will derive a NLSE by employing the reductive perturbation
method to study the MI of DAWs. So, we first introduce the stretched co-ordinates as \cite{C1,C2,C3,C4,C5,C6}
\begin{eqnarray}
&&\hspace*{-1.3cm}\xi={\epsilon}(x-v_g t),
\label{1eq:8}\\
&&\hspace*{-1.3cm}\tau={\epsilon}^2 t,
\label{1eq:9}\
\end{eqnarray}
where $v_g$ is the group speed and $\epsilon$ is a small parameter.
Then, we can write the dependent variables as \cite{C7,C8,C9,C10}
\begin{eqnarray}
&&\hspace*{-1.3cm}n_+=1+\sum_{m=1}^{\infty}\epsilon^{m}\sum_{l=-\infty}^{\infty}n_{+l}^{(m)}(\xi,\tau)~\mbox{exp}[i l(kx-\omega t)],
\label{1eq:10}\\
&&\hspace*{-1.3cm}u_+=\sum_{m=1}^{\infty}\epsilon^{m}\sum_{l=-\infty}^{\infty}u_{+l}^{(m)}(\xi,\tau)~\mbox{exp}[i l(kx-\omega t)],
\label{1eq:11}\\
&&\hspace*{-1.3cm}n_-=1+\sum_{m=1}^{\infty}\epsilon^{m}\sum_{l=-\infty}^{\infty}n_{-l}^{(m)}(\xi,\tau)~\mbox{exp}[i l(kx-\omega t)],
\label{1eq:12}\\
&&\hspace*{-1.3cm}u_-=\sum_{m=1}^{\infty}\epsilon^{m}\sum_{l=-\infty}^{\infty}u_{-l}^{(m)}(\xi,\tau)~\mbox{exp}[i l(kx-\omega t)],
\label{1eq:13}\\
&&\hspace*{-1.3cm}\phi=\sum_{m=1}^{\infty}\epsilon^{m}\sum_{l=-\infty}^{\infty}\phi^{(m)}(\xi,\tau)~\mbox{exp}[i l(kx-\omega t)],
\label{1eq:14}\
\end{eqnarray}
where $k$ ($\omega$) is real variable which represents the carrier wave number (frequency) of the solitary wave.
The derivative operators can be then written as
\begin{eqnarray}
&&\hspace*{-1.3cm}\frac{\partial}{\partial t}\rightarrow\frac{\partial}{\partial t}-\epsilon v_g \frac{\partial}
{\partial\xi}+\epsilon^2\frac{\partial}{\partial\tau},
\label{1eq:15}\\
&&\hspace*{-1.3cm}\frac{\partial}{\partial x}\rightarrow\frac{\partial}{\partial x}+\epsilon\frac{\partial}{\partial\xi}.
\label{1eq:16}
\end{eqnarray}
Now, by substituting Eqs. \eqref{1eq:8}-\eqref{1eq:16}  into  Eqs. \eqref{1eq:1}$-$\eqref{1eq:4}, and Eq. \eqref{1eq:7}, and
collecting only the terms containing $\epsilon$ for the first order ($m=1$ with $l=1$), we get the following equations
\begin{eqnarray}
&&\hspace*{-1.3cm}\omega n_{+1}^{(1)}=u_{+1}^{(1)},
\label{1eq:17}\\
&&\hspace*{-1.3cm}k\phi_1^{(1)}=\omega u_{+1}^{(1)},
\label{1eq:18}\\
&&\hspace*{-1.3cm}\omega n_{-1}^{(1)}=u_{-1}^{(1)},
\label{1eq:19}\\
&&\hspace*{-1.3cm}k\nu_1\phi_1^{(1)}=-\omega u_{-1}^{(1)},
\label{1eq:20}\\
&&\hspace*{-1.3cm}n_{+1}^{(1)}=k^2\phi_{1}^{(1)}+J_1\phi_{1}^{(1)}+\nu_2n_{-1}^{(1)},
\label{1eq:21}
\end{eqnarray}
these equations then reduce to
\begin{eqnarray}
&&\hspace*{-1.3cm}n_{+1}^{(1)}=\frac{k^2}{\omega^2}\phi_1^{(1)},
\label{1eq:22}\\
&&\hspace*{-1.3cm}u_{+1}^{(1)}=\frac{k}{\omega}\phi_1^{(1)},
\label{1eq:23}\\
&&\hspace*{-1.3cm}n_{-1}^{(1)}=-\frac{\nu_1k^2}{\omega^2}\phi_1^{(1)},
\label{1eq:24}\\
&&\hspace*{-1.3cm}u_{-1}^{(1)}=-\frac{\nu_1k}{\omega}\phi_1^{(1)}.
\label{1eq:25}\
\end{eqnarray}
The dispersion relation finally yields as
\begin{eqnarray}
&&\hspace*{-1.3cm}\omega^2=\frac{k^2(1+\nu_1\nu_2)}{k^2+J_1}.
\label{1eq:26}\
\end{eqnarray}
The second-order equations ($m=2$ with $l=1$) are given as
\begin{eqnarray}
&&\hspace*{-1.3cm}n_{+1}^{(2)}=\frac{k^2}{\omega^2}\phi_1^{(2)}+\frac{2ik(kv_g-\omega)}{\omega^3}\frac{\partial \phi_1^{(1)}}{\partial\xi},
\label{1eq:27}\\
&&\hspace*{-1.3cm}u_{+1}^{(2)}=\frac{k}{\omega}\phi_1^{(2)}+\frac{i(kv_g-\omega)}{\omega^2}
\frac{\partial \phi_1^{(1)}}{\partial\xi},
\label{1eq:28}\\
&&\hspace*{-1.3cm}n_{-1}^{(2)}=-\frac{\nu_1k^2}{\omega^2}\phi_1^{(2)}-\frac{2i\nu_1k(kv_g-\omega)}{\omega^3}
\frac{\partial \phi_1^{(1)}}{\partial\xi},
\label{1eq:29}\\
&&\hspace*{-1.3cm}u_{-1}^{(2)}=-\frac{\nu_1k}{\omega}\phi_1^{(2)}-\frac{i\nu_1(kv_g-\omega)}{\omega^2}
\frac{\partial \phi_1^{(1)}}{\partial\xi},
\label{1eq:30}\
\end{eqnarray}
with the help of compatibility condition, the group velocity of DAWs can be written as
\begin{eqnarray}
&&\hspace*{-1.3cm}v_g=\frac{\partial \omega}{\partial k}=\frac{\omega(1+\nu_1\nu_2-\omega^2)}{k(1+\nu_1\nu_2)}.
\label{1eq:31}
\end{eqnarray}
The coefficients of $\epsilon$ for $m=2$ and $l=2$ provide the second-order harmonic amplitudes
which are found to be proportional to $|\phi^{(1)}_1|^2$
\begin{eqnarray}
&&\hspace*{-1.3cm}n_{+2}^{(2)}=J_4|\phi_1^{(1)}|^2,
\label{1eq:32}\\
&&\hspace*{-1.3cm}u_{+2}^{(2)}=J_5 |\phi_1^{(1)}|^2,
\label{1eq:33}\\
&&\hspace*{-1.3cm}n_{-2}^{(2)}=J_6|\phi_1^{(1)}|^2,
\label{1eq:34}\\
&&\hspace*{-1.3cm}u_{-2}^{(2)}=J_7 |\phi_1^{(1)}|^2,
\label{1eq:35}\\
&&\hspace*{-1.3cm}\phi_{2}^{(2)}=J_8 |\phi_1^{(1)}|^2,
\label{1eq:36}\
\end{eqnarray}
where
\begin{eqnarray}
&&\hspace*{-1.3cm}J_4=\frac{3k^4+2J_8\omega^2k^2}{2\omega^4},
\nonumber\\
&&\hspace*{-1.3cm}J_5=\frac{k^3+2J_8k\omega^2}{2\omega^3},
\nonumber\\
&&\hspace*{-1.3cm}J_6=\frac{3\nu_1^2k^4-2J_8\omega^2k^2}{2\omega^4},
\nonumber\\
&&\hspace*{-1.3cm}J_7=\frac{\nu_1^2k^3-2J_8k\omega^2}{2\omega^3},
\nonumber\\
&&\hspace*{-1.3cm}J_8=\frac{2J_2\omega^4+3\nu_2\nu_1^2k^4-3k^4}{2\omega^2(k^2+\nu_1\nu_2k^2-4k^2\omega^2-J_1\omega^2)}.
\nonumber\
\end{eqnarray}
Now, we consider the expression for ($m = 3$ with $l = 0$)
and ($m = 2$ with $l = 0$) which leads to the zeroth harmonic
modes. Thus, we obtain
\begin{eqnarray}
&&\hspace*{-1.3cm}n_{+0}^{(2)}=J_9|\phi_1^{(1)}|^2,
\label{1eq:37}\\
&&\hspace*{-1.3cm}u_{+0}^{(2)}=J_{10}|\phi_1^{(1)}|^2,
\label{1eq:38}\\
&&\hspace*{-1.3cm}n_{-0}^{(2)}=J_{11}|\phi_1^{(1)}|^2,
\label{1eq:39}\\
&&\hspace*{-1.3cm}u_{-0}^{(2)}=J_{12}|\phi_1^{(1)}|^2,
\label{1eq:40}\\
&&\hspace*{-1.3cm}\phi_0^{(2)}=J_{13}|\phi_1^{(1)}|^2,
\label{1eq:41}\
\end{eqnarray}
where
\begin{eqnarray}
&&\hspace*{-1.3cm}J_9=\frac{\omega k^2+J_{13}\omega^3+2v_gk^3}{v_g^2\omega^3},
\nonumber\\
&&\hspace*{-1.3cm}J_{10}=\frac{k^2+J_{13}\omega^2}{v_g\omega^2},
\nonumber\\
&&\hspace*{-1.3cm}J_{11}=\frac{\omega\nu_1^2 k^2+2v_g\nu_1^2k^3-J_{13}\nu_1\omega^3}{v_g^2\omega^3},
\nonumber\\
&&\hspace*{-1.3cm}J_{12}=\frac{\nu_1^2 k^2-J_{13}\nu_1\omega^2}{v_g\omega^2},
\nonumber\\
&&\hspace*{-1.3cm}J_{13}=\frac{2v_g(J_2v_g\omega_3-k^3+\nu_2\nu_1^2k^3)-\omega k^2(1-\nu_2\nu_1^2)}{\omega^3+\nu_1\nu_2\omega^3-J_1\omega^3v_g^2}.
\nonumber\
\end{eqnarray}
Finally, the third harmonic modes (m = 3) and (l = 1),
with the help of Eqs. \eqref{1eq:22}$-$\eqref{1eq:41}, give a set of equations which
can be reduced to the following NLSE:
\begin{eqnarray}
&&\hspace*{-1.3cm}i\frac{\partial\Phi}{\partial\tau}+P\frac{\partial^2\Phi}
{\partial\xi^2}+Q|\Phi|^2\Phi=0,
\label{1eq:42}
\end{eqnarray}
where $\Phi=\phi_1^{(1)}$ for simplicity. In Eq. \eqref{1eq:42}, $P$ is the dispersion coefficient and can be written as
\begin{eqnarray}
&&\hspace*{-1.3cm}P=\frac{3v_g(kv_g-\omega)}{2k\omega},
\nonumber\
\end{eqnarray}
and $Q$ is the nonlinear coefficient and  can be written as
\begin{eqnarray}
&&\hspace*{-1.3cm}Q=\frac{3J_3\omega^3+2J_2\omega^3(J_8+J_{13})-F}{2k^2(1+\nu_1\nu_2)},
\nonumber\
\end{eqnarray}
where
\begin{eqnarray}
&&\hspace*{-1.3cm}F=\omega k^2(J_4+J_9)+\nu_1\nu_2\omega k^2(J_6+J_{11})
\nonumber\\
&&\hspace*{-0.8cm}+2k^3(J_5+J_{10})+2\nu_1\nu_2k^3(J_7+J_{12}).
\nonumber\
\end{eqnarray}
It may be essential to note that the values of $P$ and $Q$ depend on the physical parameters
of the system and any change in these parameters will directly affect the nonlinearity and dispersion
properties of the EDDPM.
\begin{figure}
\centering
\includegraphics[width=80mm]{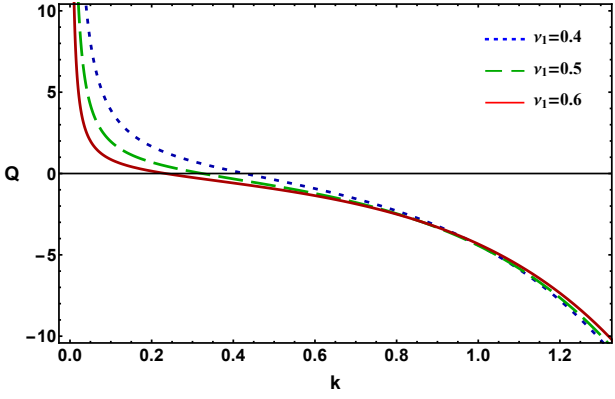}
\caption{Plot of $Q$ vs $k$ for different values of $\nu_1$ when $\nu_2=2.5$ and $q=1.7$.}
 \label{1Fig:F2}
\end{figure}
\begin{figure}
\centering
\includegraphics[width=80mm]{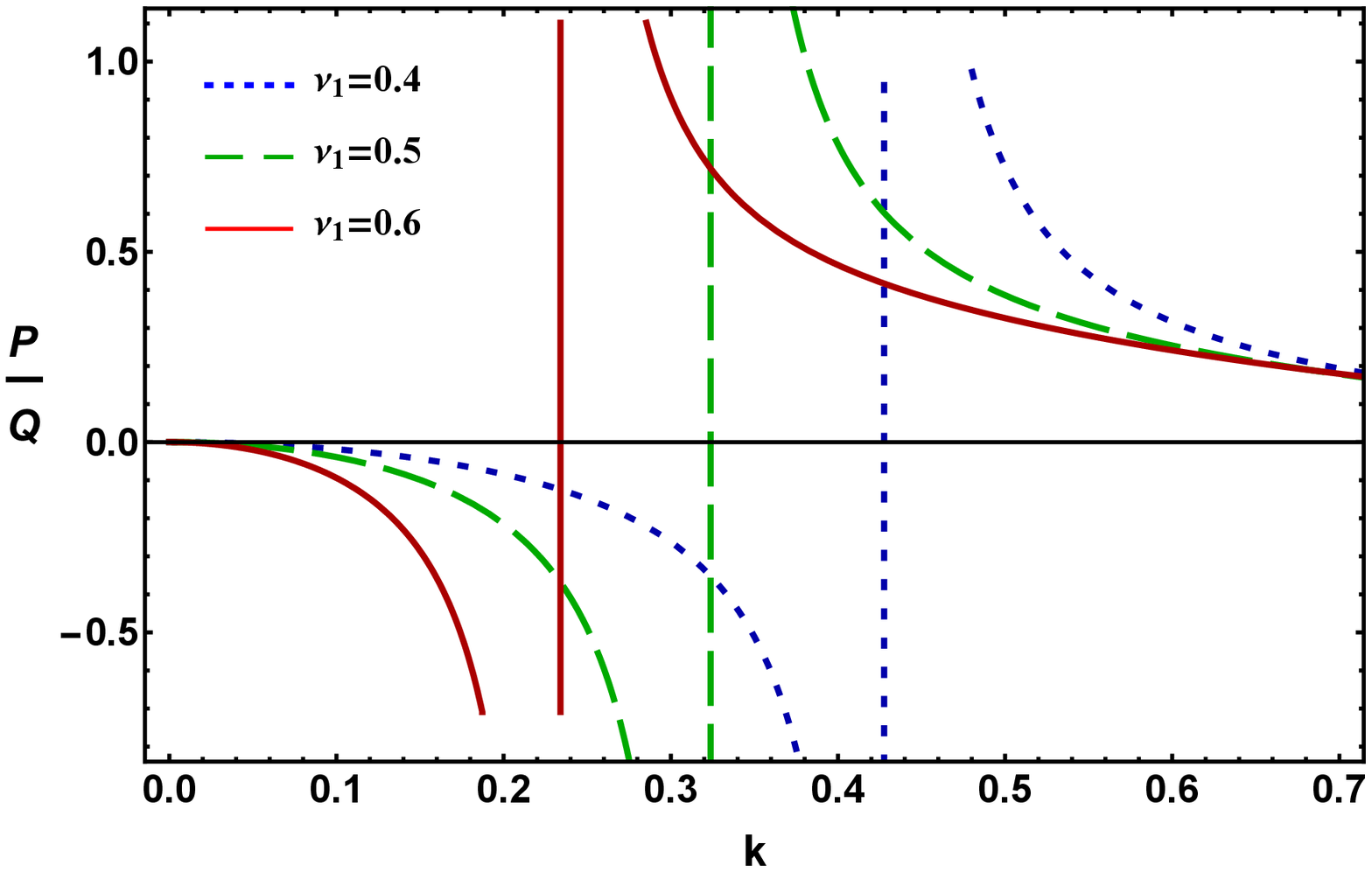}
\caption{Plot of $P/Q$ vs $k$ for different values of $\nu_1$ when $\nu_2=2.5$ and $q=1.7$.}
\label{1Fig:F3}
\end{figure}
\begin{figure}
\centering
\includegraphics[width=80mm]{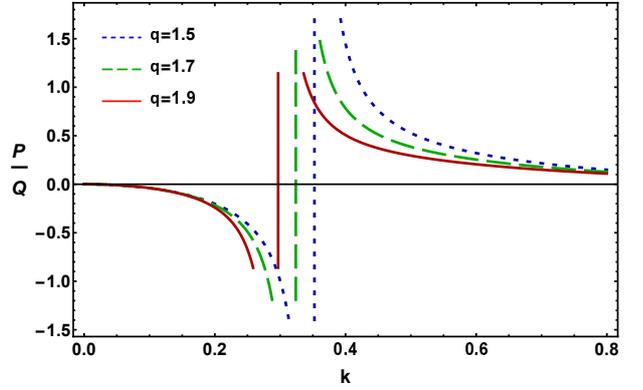}
\caption{Plot of $P/Q$ vs $k$ for $q$=positive when $\nu_1=0.5$ and $\nu_2=2.5$.}
 \label{1Fig:F4}
\end{figure}
\begin{figure}
\centering
\includegraphics[width=80mm]{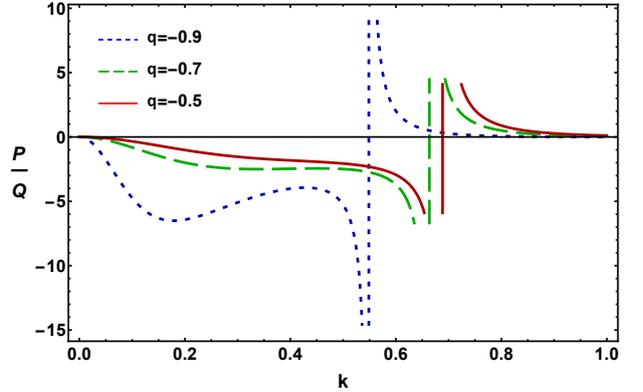}
\caption{Plot of $P/Q$ vs $k$ for $q$=negative when $\nu_1=0.5$ and $\nu_2=2.5$.}
\label{1Fig:F5}
\end{figure}
\begin{figure}
\centering
\includegraphics[width=80mm]{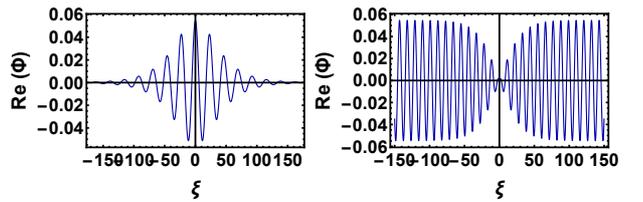}
\caption{Plot of Re$(\Phi)$ vs $\xi$ for bright envelope soliton (left panel) when $k=0.4$ and
dark envelope soliton (right panel) when $k=0.1$ . Other parameters are $\nu_1=0.5$,
$\nu_2=2.5$, $\tau=0$, $\psi_0=0.007$, $U=0.3$, $\Omega_0=0.2$, and $q=1.7$.}
 \label{1Fig:F6}
\end{figure}
\begin{figure}
\centering
\includegraphics[width=80mm]{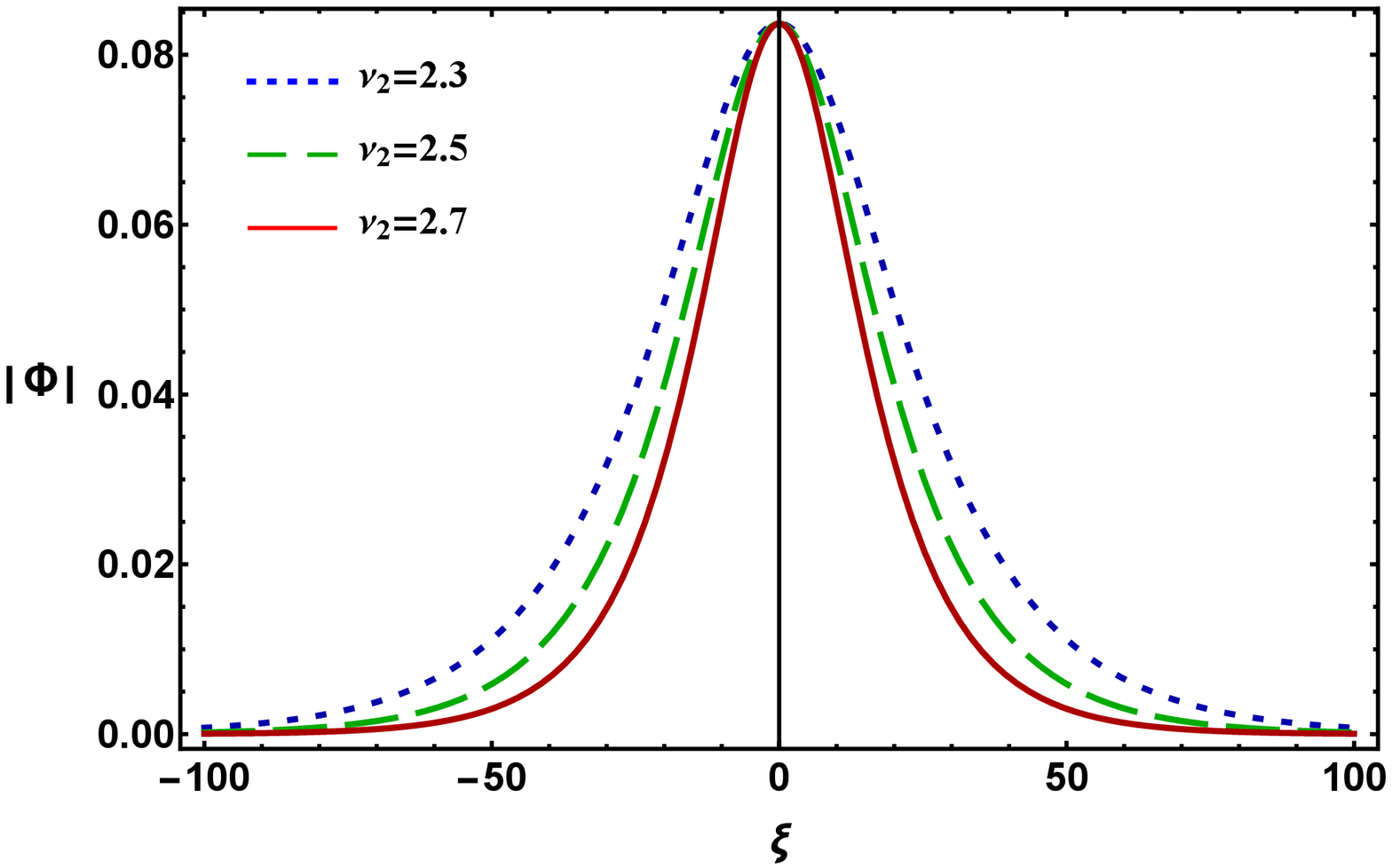}
\caption{Plot of $|\Phi|$ vs $\xi$ for different values of $\nu_2$ when $q=1.7$, $\nu_1=0.5$,
$\tau=0$, $\psi_0=0.007$, $U=0.3$, $\Omega_0=0.2$, and $k=0.4$.}
 \label{1Fig:F7}
\end{figure}
\begin{figure}[t!]
\centering
\includegraphics[width=80mm]{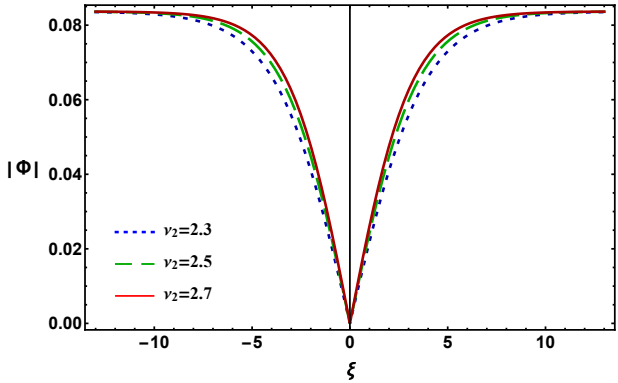}
\caption{Plot of $|\Phi|$ vs $\xi$ for different values of $\nu_2$ when $q=1.7$, $\nu_1=0.5$,
$\tau=0$, $\psi_0=0.007$, $U=0.3$, $\Omega_0=0.2$, and $k=0.1$.}
\label{1Fig:F8}
\end{figure}
\section{Modulational instability}
\label{1sec:Modulational instability}
The stable and unstable parametric regimes of DAWs are
organised by the sign of $P$ and $Q$ of Eq. \eqref{1eq:42} \cite{C11,C12,C13,Fedele2002}.
When $P$ and $Q$ have the same sign (i.e., $P/Q > 0$),
the evolution of DAWs amplitude is modulationally
unstable in the presence of external perturbations.
On the other hand, when $P$ and $Q$ have opposite signs
(i.e., $P/Q < 0$), the DAWs are modulationally stable in
the presence of external perturbations. The plot of $P/Q$
against $k$ yields stable and unstable parametric regimes
of the DAWs. The point, at which the transition of $P/Q$
curve intersects with the $k$-axis, is known as the threshold
or critical wave number $k~(= k_c)$ \cite{C11,C12,C13,Fedele2002}.

We have depicted the variation of $P$ with $k$ for different values of $\nu_1$ in Fig. \ref{1Fig:F1}
and this depiction indicates that (a) the $P$ is always negative for any
values of $k$ in the EDDPM; (b) the absolute value of $P$ increases (decreases)
with an increase in the value of positive (negative) dust mass for their constant
charge state (via $\nu_1$).

Figure \ref{1Fig:F2} indicates how the
MI of the DAWs is directly organized by the nonlinear  coefficient $Q$ of Eq. \eqref{1eq:40},
and it is obvious from this figure that (a) $Q$ can be positive/negative according to the
value of $k$ and other plasma parameters; (b) $Q$ has positive (negative) value corresponding to the
small (large) values of $k$; (c) as $Q$ has a combination of both positive and negative values
instead of only negative value like $P$ then the MI picture of the DAWs can be directly
organized by flipping the sign of $Q$ as well as $P/Q$.

The criteria of the formation of dark envelope solitons associated with stable region
(i.e., $P/Q<0$) of DAWs as well as bright envelope solitons associated with unstable
region (i.e., $P/Q>0$) of DAWs can be shown in Fig. \ref{1Fig:F3} in which the variation
of $P/Q$ with $k$ for different values of $\nu_1$ is depicted. It can be seen from this figure
that (a) an increase in the value of negative (positive) dust mass enhances (reduces)
the modulationally stable parametric regime for a fixed value of $Z_-$ and $Z_+$;
(b) the modulationally stable parametric regime decreases (increases)
with $Z_-$ ($Z_+$) for a fixed value of positive and negative dust mass.

The effects of both sub-extensivity and super-extensivity of ions on
the stable and unstable parametric regimes of DAWs can be observed from  Figs. \ref{1Fig:F4}
and \ref{1Fig:F5}, respectively,  and it is obvious from these figures that (a) for sub-extensive limit
(i.e., $q>1$), the $k_c$ as well as the stable parametric regime (i.e., $P/Q<0$) of DAWs
increases with an decrease in the value of $q$; (b) the electrostatic dark envelope solitons
associated with DAWs can generate for small values of $k$ (i.e., $k<k_c$) while the
electrostatic bright envelope solitons associated with DAWs can generate for large values
of $k$ (i.e., $k>k_c$); (c) on the other hand, for super-extensive limit (i.e., $-1<q<1$), the
$k_c$ as well as the stable parametric regime (i.e., $P/Q<0$) of DAWs increases with an
increase in the value of $q$.
\section{Envelope solitons}
\label{1sec:Envelope solitons}
The bright (when $P/Q>0$) and dark (when $P/Q<0$) envelope
solitonic solutions, respectively, can be written as \cite{Jahan2019}
\begin{eqnarray}
&&\hspace*{-1.3cm}\Phi(\xi,\tau)=\left[\psi_0~\mbox{sech}^2 \left(\frac{\xi-U\tau}{W}\right)\right]^\frac{1}{2}
\nonumber\\
&&\hspace*{-0.01cm}\times \exp \left[\frac{i}{2P}\left\{U\xi+\left(\Omega_0-\frac{U^2}{2}\right)\tau \right\}\right],
\label{1eq:43}\\
&&\hspace*{-1.3cm}\Phi(\xi,\tau)=\left[\psi_{0}~\mbox{tanh}^2 \left(\frac{\xi-U\tau}{W}\right)\right]^\frac{1}{2}
\nonumber\\
&&\hspace*{-0.01cm}\times \exp \left[\frac{i}{2P}\left\{U\xi-\left(\frac{U^2}{2}-2 P Q \psi_{0}\right)\tau \right\}\right],
\label{1eq:44}
\end{eqnarray}
where $\psi_0$ is the amplitude of localized pulse for both bright and dark
envelope soliton, $U$ is the propagation speed of the localized pulse, $W$ is the soliton width, and
$\Omega_0$ is the oscillating frequency at $U=0$. The soliton width $W$ and the maximum amplitude  $\psi_0$
are related by $W=\sqrt{2\mid P/Q\mid/\psi_0}$. We have depicted bright and dark envelope
solitons profile in Fig. \ref{1Fig:F6}, and also numerically analyzed the effect of the physical parameter
$\nu_2$ on the profile of bright and dark envelope solitons in Figs. \ref{1Fig:F7} and \ref{1Fig:F8}, respectively.
It is clear from these figures that the increase in the value of $\nu_2$ only causes to
change the width of both bright and dark envelope solitons associated with DAWs but the magnitude of the amplitude of
both bright and dark envelope solitons associated with DAWs does not change.
\section{Conclusion}
\label{1sec:Conclusion}
In our article, a NLSE has been employed to study the MI and associated nonlinear features
of DAWs in an unmagnetized EDDPM comprising OPDGs and $q$-distributed positive ions.
The physical parameters of the plasma system rigorously control the balance between the nonlinearity and
dispersion of the medium hence set the criterion for MI of DAWs as well as the formation
of both bright and dark envelope solitons. The sub-extensive (super-extensive)
ions help to decrease (increase) the stable region of DAWs. An addition of charge
in the positive (negative) DGs enhances the stability (unstability) region.
The width of the solitons is changed by the variation of the plasma parameters while the amplitude
of the solitons is unchanged. Finally, we hope that our investigation is useful in understanding
the nonlinear behavior of DAWs and the propagation of envelope solitons
in various space plasma environments, viz., cometary tail \cite{Hossen2016a},
Martian atmosphere \cite{Hossen2016b}, interstellar clouds \cite{Hossen2017},
and laser-matter interaction \cite{Shahmansouri2013a}, etc.


\begin{thebibliography}{99}

\bibitem{Sahu2012a} B. Sahu and M. Tribeche, Astrophys. Space Sci. \textbf{338}, 259 (2012).

\bibitem{Hossen2016a} M.M. Hossen, M.S. Alam, \textit{et al.}, Eur. Phys. J. D \textbf{70}, 252 (2016).

\bibitem{Hossen2016b} M.M. Hossen, M.S. Alam, \textit{et al.}, Phys. Plasmas \textbf{23}, 023703 (2016).

\bibitem{Hossen2017} M.M. Hossen, L. Nahar, \textit{et al.}, High Energ. Dens. Phys. \textbf{24}, 9 (2017).

\bibitem{Kopnin2007} S.I. Kopnin, S.I. Popel, and M.Y. Yu, Plasma Phys. Rep. \textbf{33}, 289 (2007).

\bibitem{Shahmansouri2013a} M. Shahmansouri and H. Alinejad, Phys. Plasmas \textbf{20}, 033704 (2013).

\bibitem{Ferdousi2017} M. Ferdousi, S. Sultana, \textit{et al.}, Eur. Phys. J. D. \textbf{71}, 102 (2017).

\bibitem{Ferdousi2015} M. Ferdousi, M.R. Miah, \textit{et al.}, Astrophys. Space Sci. \textbf{43}, 360 (2015).

\bibitem{Sahu2012b} B. Sahu and M. Tribeche, Astrophys. Space Sci. \textbf{341}, 573 (2012).

\bibitem{Kopnin2009} S.I. Kopnin, S.I. Popel, and M.Y. Yu, Phys. Plasmas \textbf{16}, 063705 (2009).

\bibitem{Popel1994} S. I. Popel and M. Y. Yu, Phys. Rev. E \textbf{50}, 3060 (1994).

\bibitem{Rao1990} N.N. Rao, P.K. Shukla, and M.Y. Yu, Planet. Space Sci. \textbf{38}, 543 (1990).

\bibitem{Shukla1992} P.K. Shukla and V.P. Silin, Phys. Scr. \textbf{45}, 508 (1992).

\bibitem{Tsallis1988} C. Tsallis, J. Stat. Phys. \textbf{52}, 479 (1988).

\bibitem{Renyi1955} A. Renyi, Acta Math. Acad. Sci. Hung.  \textbf{6}, 285 (1955).

\bibitem{Saha2014} A. Saha and P. Chatterjee, Astrophys. Space Sci. \textbf{353}, 169 (2014).

\bibitem{Zaghbeer2014a} S.K. Zaghbeer, H.H. Salah, \textit{et al.}, J. Plasma Phys. \textbf{80}, 517 (2014).

\bibitem{Zaghbeer2014b} S.K. Zaghbeer, H.H. Salah, \textit{et al.}, Astrophys. Space Sci. \textbf{353}, 493 (2014).

\bibitem{Roy2014} K. Roy, P. Chatterjee, \textit{et al.}, Astrophys. Space Sci. \textbf{350}, 599 (2014).

\bibitem{Bacha2012} M. Bacha and M. Tribeche, Astrophys. Space Sci. \textbf{337}, 253 (2012).

\bibitem{Eastman1998} T.E. Eastman, \textit{et al.},  \textit{et. al.}, Geophys. Res. Lett. \textbf{103}, 23503 (1998).

\bibitem{Pavlos2011} G.P. Pavlos, A.C. Iliopoulos, \textit{et al.}, Physica A \textbf{390}, 2819 (2011).

\bibitem{Liu1994} J.M. Liu, J.S. De Groot, \textit{et al.}, Phys. Rev. Lett. \textbf{72}, 2717 (1994).

\bibitem{Jahan2019} S. Jahan, \textit{et al.}, Commun. Theor. Phys. \textbf{71}, 327 (2019).

\bibitem{Saini2008} N.S. Saini and I. Kourakis, Phys. Plasmas \textbf{15}, 123701 (2008).

\bibitem{El-Taibany2006} W.F. El-Taibany and I. Kourakis, Phys. Plasmas \textbf{13}, 062302 (2006).

\bibitem{Demiray2019} H. Demiray and A. Abdikian, Chaos Solitons Fractals, \textbf{121}, 50 (2019).

\bibitem{Moslem2011} W. M. Moslem, R. Sabry, \textit{et al.}, Phys. Rev. E  \textbf{84}, 066402 (2011).

\bibitem{Vladimirov1995} S. V. Vladimirov, V. N. Tsytovich, \textit{et al.}, \textit{Modulational Interactions in Plasmas} (Kluwer Academic, Dordrecht, 1995).

\bibitem{Gailitis1964} A. K. Gailitis, Ph.D. Thesis (Lebedev Physical Institute, USSR Acad. Sci., Moscow, 1964).

\bibitem{Gailitis1965} A. K. Gailitis, Izv. AN Latv. SSR: Phys. Tech. Nauki \textbf{4},
13 (1965).

\bibitem{Vedenov1965} A. A. Vedenov and L. I. Rudakov, Sov. Phys. Doklady \textbf{9}, 1073 (1965).

\bibitem{Tagare1997} S. G. Tagare, Phys. Plasmas \textbf{04}, 3167 (1997).

\bibitem{Ghai2018} Y. Ghai, N. Kaur, \textit{et al.}, Plasma Sci. Technol. \textbf{20}, 074005 (2018).

\bibitem{Shahmansouri2012} M. Shahmansouri and M. Tribeche, Astrophys. Space Sci. \textbf{342}, 87 (2012).

\bibitem{Shahmansouri2013b} M. Shahmansouri, Iran. J. Sci. Technol. \textbf{37}(A3), 285 (2013).

\bibitem{El-Tantawy2012} S. A. El- Tantawy, Astrophys. Space Sci. \textbf{337}, 209 (2012).

\bibitem{C1} N.A. Chowdhury, \textit{et al.}, Chaos \textbf{27}, 093105 (2017).

\bibitem{C2} S.K. Paul, \textit{et al.}, Pramana-J Phys \textbf{94} (2020) 58.

\bibitem{C3} M.H. Rahman, \textit{et al.}, Chinese J. Phys. \textbf{56}, 2061 (2018).

\bibitem{C4} M.H. Rahman, \textit{et al.}, Phys. Plasmas \textbf{25}, 102118 (2018).

\bibitem{C5} N.A. Chowdhury, \textit{et al.}, Vacuum \textbf{147}, 31 (2018).

\bibitem{C6} N.A. Chowdhury, \textit{et al.}, Contrib. Plasma Phys. \textbf{58}, 870 (2018).

\bibitem{C7} N.Ahmed, \textit{et al.}, Chaos \textbf{28}, 123107 (2018).

\bibitem{C8} N.A. Chowdhury, \textit{et al.}, Plasma Phys. Rep. \textbf{45}, 459 (2019).

\bibitem{C9} M. Hassan, \textit{et al.}, Commun. Theor. Phys. \textbf{71}, 1017 (2019).

\bibitem{C10} R.K. Shikha, \textit{et al.}, Eur. Phys. J. D \textbf{73}, 177 (2019).

\bibitem{C11} S. Jahan, \textit{et al.}, Plasma Phys. Rep. \textbf{46}, 90 (2020) .

\bibitem{C12} N.A. Chowdhury, \textit{et al.}, Phys. plasmas \textbf{24}, 113701 (2017).

\bibitem{C13} T.I. Rajib, \textit{et al.}, Phys. plasmas \textbf{26} 123701 (2019) .

\bibitem{Fedele2002} R. Fedele, Phys. Scr. \textbf{65}, 502 (2002).

\end{thebibliography}
\end{document}